\documentclass[]{vldb}

\usepackage{algorithm}
\usepackage[noend]{algpseudocode}
\usepackage{amsmath}
\usepackage{amssymb}
\usepackage{booktabs} 
\usepackage{color}
\usepackage{multirow}
\usepackage{balance}
\usepackage{subcaption}
\usepackage[geometry]{ifsym}
\usepackage{enumitem}
\usepackage{url}

\usepackage{amsthm}

\newcommand{\sys}{Quickstep}
\newcommand{\SYS}{QUICKSTEP}
\newcommand{\wo}{work order}

\newcommand{\todo}[1]{}
\renewcommand{\todo}[1]{{\color{red} TODO: {#1}}}

\theoremstyle{plain}

\vldbTitle{To pipeline, or not to pipeline, that is the question}
\subtitle{Clarifying intra-operator data transfer mechanisms for in-memory data systems}
\vldbAuthors{Harshad Deshmukh, Bruhathi Sundarmurthy, Jignesh M. Patel}
\vldbDOI{https://doi.org/10.14778/xxxxxxx.xxxxxxx}
\vldbVolume{12}
\vldbNumber{xxx}
\vldbYear{2019}

\begin{document}
\title{To pipeline, or not to pipeline, that is the question}
\subtitle{Clarifying intra-operator data transfer mechanisms for in-memory data systems}
\numberofauthors{1}
\author{Harshad Deshmukh\thanks{Work done while at UW-Madison}, Bruhathi Sundarmurthy{$^*$}, Jignesh M. Patel{$\,^{\dag}$}\\
\affaddr{$^*$Google, $\,^{\dag}$University of Wisconsin - Madison}\\
\affaddr{\{harshad, bruhathi, jignesh\}@cs.wisc.edu}
}
\maketitle
	
	\begin{abstract}
In designing query processing primitives, a crucial design choice is the method for data transfer 
between two operators in a query plan. As we were considering this critical design mechanism for an in-memory 
database system that we are building, we quickly realized that (surprisingly) there isn't a clear 
definition of this concept. Papers are full or ad hoc use of terms like pipelining and blocking, 
but as these terms are not crisply defined, it is hard to fully understand the results
attributed to these concepts. To address this limitation, we introduce a clear terminology for how 
to think about data transfer between operators in a query pipeline. We show that there isn't a 
clear definition of pipelining and blocking, and that there is a full spectrum of techniques based on 
a simple concept called unit-of-transfer. Next, we develop an analytical model for inter-operator 
communication, and highlight the key parameters that impact performance (for in-memory database settings). 
Armed with this model, we then apply it to the system we are designing and highlight the 
insights we gathered from this exercise. We find that the gap between pipelining and non-pipelining 
query execution, w.r.t. key factors such as performance and memory footprint is quite narrow, 
and thus system designers should likely rethink the notion of pipelining vs. blocking for in-memory 
database systems.

\end{abstract}
	
	%
	%
	
%

\section{Introduction}
A fundamental consideration in query processing design is the mechanism for communicating
data between one operator and another, such as a select operator feeding to an
aggregate operator, or a select operator feeding to a probe operator to evaluate a
hash join. Typically the source operator is called the \textit{producer} and the
destination is called the \textit{consumer}.
There are two broad camps for \textit{intra-operator communication} methods, in both traditional disk-based and newer in-memory systems. These two camps sharply distinguish themselves based on the data transfer method between producers and consumer. These camps are pipelining (e.g.~\cite{vectorwise, hyper}) and blocking (e.g.~\cite{monetdb, DBLP:conf/osdi/DeanG04}).

\begin{figure}[t]
	\centering 
	\includegraphics[width=\columnwidth]{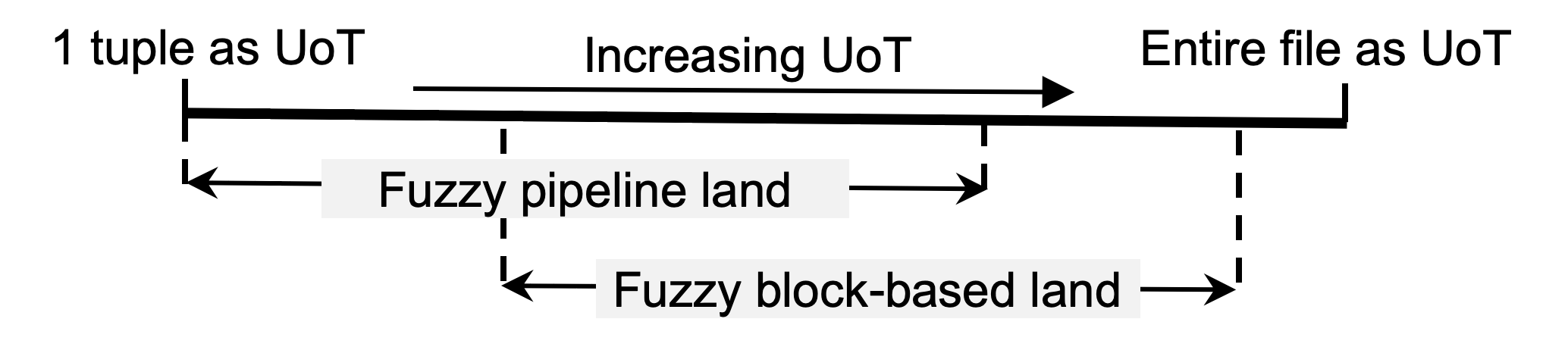}
	\caption{\textbf{Unit of Transfer (UoT)}}
	\label{fig:uot}
\end{figure}

Understanding the implication of choosing one method over the other is non-trivial since there are varied definitions of what comprises pipelining or blocking. For example, in~\cite{DBLP:journals/pvldb/KerstenLKNPB18}, the definition of pipeline leans towards ``a tuple being processed should be present in the register''. Vectorwise~\cite{vectorwise} departed  from the traditional tuple-at-a-time processing model and proposed hyper-pipelining query execution~\cite{hyper-pipelining} using batches (or vectors) of tuples. On the other hand, disk-based systems~\cite{DBLP:conf/sigmod/DeWittGS88, gerber1986phd,DBLP:conf/vldb/SchneiderD90} define pipelining as ``tuples should be successively processed without having to be sent to the disk in between''. 

From these examples, we observe that the line between pipelining and blocking is fuzzy and depends on the batch size of data transfer. The first key contribution of this paper is to highlight that there is no crisp definition of pipelining or blocking and that is a huge source of confusion. It is hard to understand results either for or against these mechanisms without a crisp definition. In this paper we introduce the term \emph{unit-of-transfer} (UoT) to clarify these mechanisms. This simple concept is graphically depicted in Figure~\ref{fig:uot}.

With this terminology we can see that the granularity of inter-operator transfer mechanisms is really a spectrum; different systems are designed to support different UoT values.
At one end of the spectrum, a tuple can be the UoT~\cite{hyper}, whereas at the other end of the spectrum, the whole table (or the whole file or the whole intermediate result) can be the UoT~\cite{DBLP:conf/osdi/DeanG04}. Many systems such as MonetDB~\cite{monetdb} and \sys{}, which produce batches of tuples as output, fall somewhere in between the two extremes.

We point out an immediate benefit of introducing the notion of
UoT -- it implicitly addresses the confusion about where data should reside in the memory hierarchy for the data transfer mechanism to be called `pipelining' or otherwise. For instance, if the UoT is very small,
chances are that it is resident in registers, or if the UoT is too big to
fit in the memory, it may be forced to disk/persistent storage. As traditional
disk-based systems are affected by expensive disk I/O operations,
we can say that their UoT is a batch of tuples that are main-memory
resident.

Next, with a new clear terminology of UoT for data transfer mechanisms, we propose a model to study the implication
of changing UoT for in-memory systems. Here, we enumerate key factors that are crucial to the model and their interaction.
The factors are parallelism, block size, storage format, query structure, and hardware characteristics. A combination of these factors jointly impacts the performance of a query when there is no I/O bottleneck. 
The collective space of combinations for these dimensions is very large, and prior work has largely looked at individual dimensions and studied their impact on overall query execution. 

The analytical model helps understand the factors that are important and the performance implication of changing values for these factors. It presents system designers and practitioners with a tool to analyze the impact of UoT and other dimensions on query processing performance. 


Then, as a case in point, we apply the proposed model to a specific system, \sys{}~\cite{quickstep-system} (the system background is described in Section~\ref{sec:sys-background}), and study the implication on performance. We obtain the following interesting insights for \sys{} (Note we are not claiming that the following insights apply to \textit{all} in-memory systems; we are only sharing the insights we obtained about \sys{}.):
\begin{itemize} 
	\item We observe that the performance of queries depends on many dimensions like parallelism, block size, storage format, query plan structure and hardware characteristics like prefetching. 
	\item We compare the memory requirements of query execution with changing UoT and observe that for TPC-H queries, the average memory overhead can be less than 4\% of the base table.
	\item For smaller block sizes, using a smaller UoT results in higher performance compared to using a bigger UoT. As block sizes increase, the UoT does not have much impact. This was a surprising insight for us given the amount of attention pipelined query processing has received in the past. 
\end{itemize}

Our paper is organized as follows: In Sections~\ref{sec:pipe-background} and ~\ref{sec:sys-background} we cover essential  background and discuss related work.
We discuss the dimensions associated with this study in Section~\ref{sec:dimensions}, and present our analytical model in Section~\ref{sec:model}. We compare memory footprints of strategies with different UoTs in Section~\ref{sec:memory}.
In Section~\ref{sec:experiments}, we present our experimental evaluations.
Finally, Section~\ref{sec:conclusions} contains our concluding remarks.

\section{Background and Related Work}\label{sec:pipe-background}
In this section we describe the basics of data transfer mechanisms for query processing which we use to set the discussion for the rest of the paper, and discuss related work.

\paragraph*{\textbf{Data-transfer mechanisms}} 
Since most related works in this area use the word `pipeline', we will first describe a `pipeline' so that we can refer to previous work using their own terminology. 
However, we emphasize that a pipeline, in fact, is one of the many possible data transfer mechanisms.

A minimal pipeline in a query plan consists of two operators: A \textit{producer} operator and a \textit{consumer} operator. 
The output of the producer operator can be passed (or \textit{streamed}) to the consumer operator, even when the work for the producer operator has not finished. 
An example of such a simple pipeline is a query plan with a logical select operator feeding into a logical hash-based join operator. Here, one simple (physical) pipeline is a select operation (on the probe side) feeding into a ``probe'' operation.

Deeper pipelines may consist of more than two operators, such that any two adjacent operators can form a producer and consumer pair.
Data from the original producer operator can be passed all the way to the last operator in the pipeline. 
An example of a deep pipeline is a left-deep plan for a multi-way hash join query plan, with all hash tables on the build side being resident in memory.

%
There are two key aspects about pipelining, or in general data transfer mechanisms: \textit{Materialization} (or the lack of) and \textit{eager} execution of consumer operator on the output of the producer operator. 
Different systems may vary in the representation that is used for the temporary data, which is the output of a producer operator. 
Systems such as MonetDB~\cite{monetdb} and \sys{} 
that employ a block-style query processing model fully materialize the output. 
Vectorwise~\cite{vectorwise} has a compact representation of the intermediate output and does not fully materialize the output.
Systems such as Hyper~\cite{hyper} and LegoBase~\cite{legobase} generate compiled code for the full pipeline. Therefore, they do not need an explicit representation for the temporary data (the code generator picks the internal representation).

\paragraph*{\textbf{Prior Work}} Pipelining in database systems has been studied extensively. 
Most recently Wang et al.~\cite{wang2016elastic} proposed an iterator model for pipelining in in-memory database clusters.
Their key idea is to provide flexibility in the traditional iterator through operations such as expand and shrink.
Neumann~\cite{DBLP:journals/pvldb/Neumann11} proposed compilation techniques for query plans, which is used by Hyper~\cite{hyper, morsel}. 
As discussed earlier, query compilation is one of the techniques for realizing pipelining in a query plan. 
Vectorwise~\cite{vectorwise} pioneered the vectorized query processing model through the hyper-pipelining query execution~\cite{hyper-pipelining}.
Departing from the traditional tuple-at-a-time processing model, Vectorwise used batches (or vectors) of tuples. 
These batches, potentially amenable to SIMD instructions help improve Vectorwise's perforance over its predecessor MonetDB~\cite{monetdb}.

Kersten et al. in their work on query compilation and vectorization~\cite{DBLP:journals/pvldb/KerstenLKNPB18} provide a comprehensive summary 
of pipelining in many systems, from systems as old as System-R~\cite{DBLP:journals/csur/Kim79} to modern systems like Hekaton~\cite{DBLP:journals/debu/FreedmanIL14}.
The authors describe two approaches to pipelining, namely the pull (\textit{next} interface) and push (\textit{producer/consumer} interface) model. 
\sys{} uses the push model of pipelining. 

Menon et al. proposed \textit{Relaxed Operator Fusion} model~\cite{rof} to bring together techniques like compilation, vectorization and software prefetching in a single query processing engine Peloton~\cite{pelotondb}. 
Funke et al. showed~\cite{DBLP:conf/sigmod/FunkeBNMT18} how pipelined query processing can work with query compilation and GPU accelerated database systems.

There is large body of prior work on the effect of storage format and page layouts on query performance~\cite{quickstep-storage, DBLP:journals/vldb/AilamakiDH02, DBLP:conf/vldb/HankinsP03, DBLP:journals/pvldb/GrundKPZCM10, DBLP:conf/sigmod/AbadiMH08}.
In our work, we focus on using row store and column store format for the comparison between various pipelining strategies. 

Incorporating parallelism for query execution within single node database deployment has been an active area of study since the prevalence of multi-core computing, which is exemplified by many modern systems ~\cite{quickstep-system, morsel, vectorwise, hyrise-website}. 

Liu and Rundensteiner~\cite{DBLP:conf/vldb/LiuR05} studied pipelined parallelism in bushy plans and propose alternatives to maximal pipeline processing. 
Their work focuses on optimizing query plans in the distributed execution environment with limited memory per node.
Our work differs from them in multiple aspects: We focus on single node in-memory query execution with large intra-operator parallelism. 
We focus on the query scheduler phase, which comes after the optimal query plan has been generated by the optimizer.

Zhu et al. proposed \textit{look ahead techniques} to increase robustness of query plans~\cite{DBLP:journals/pvldb/ZhuPSP17} in the Quickstep system.
Their key idea is to minimize the data that passes from the producer operator to the consumer operator in a pipeline by employing a sequence of bloom filters.

Pipelines in many TPC-H queries begin with filtering a large table (e.g. lineitem).
Researchers have looked at sharing this large amount of work across multiple queries~\cite{DBLP:conf/sigmod/HarizopoulosSA05, DBLP:conf/vldb/ZukowskiHNB07}.
Scan sharing has shown significant improvements in query performance, especially in the disk setting.

Many commercial systems including SQL Server~\cite{DBLP:conf/sigmod/LeeKNDCEEKWPNLN16}, Oracle~\cite{DBLP:conf/sigmod/CruanesDG04}, IBM DB2~\cite{DBLP:journals/pvldb/RamanABCKKLLLLMMPSSSSZ13},  Snowflake~\cite{DBLP:conf/sigmod/DagevilleCZAABC16} make use of pipelining.
SQL Server's query progress estimation techniques rely on pipelines within a query plan~\cite{DBLP:conf/sigmod/LeeKNDCEEKWPNLN16, DBLP:conf/sigmod/ChaudhuriNR04}.

In distributed settings, systems like MapReduce~\cite{DBLP:conf/osdi/DeanG04} and Dryad~\cite{DBLP:conf/eurosys/IsardBYBF07} favor reliability over pipelining, thus they materialize the intermediate data during a job. 
A recent proposal called \textit{Bubble Execution}~\cite{DBLP:journals/pvldb/YinSLELFBSD18} involves
breaking a query execution plan in \textit{bubbles} such that data can be streamed within a bubble, while still offering reliability guarantees.

\section{\SYS{} Background}\label{sec:sys-background}
In this section we provide a brief background of \sys{}~\cite{quickstep-system} and its implementation of different data transfer mechanism strategies. 
We introduce the system earlier on to facilitate the subsequent discussion on various dimensions and the experimental results.

\sys{} aims for high performance for in-memory analytic workloads on a single node.
One of the techniques used by \sys{} to get high performance is large intra-operator parallelism. 
\sys{} uses a cost-based optimizer to generate query plans. 
Joins in \sys{} use non-partitioned hash-based implementation. 
The operators in \sys{} process a batch of input tuples, rather than one tuple at a time. 
Prior work~\cite{hyper-pipelining} has shown that the vectorized style processing outperforms tuple-at-a-time processing technique.

\sys{} uses an abstraction called \textit{\wo{}s}, which represents the relational operator logic that needs to be executed on a specified input.
The work done for a query is broken up in a series of \wo{}s.
These \wo{}s can be executed independently and in parallel.

There are two kinds of threads in \sys{} -- \textit{workers} and \textit{scheduler}.
Worker threads execute \wo{}s.
Once assigned a \wo{}, the worker thread executes it until its completion. 
A single scheduler thread coordinates the execution of \wo{}s, including \wo{} dispatch and \wo{} progress monitoring.

\subsection{Managing Storage in \sys{}}
\sys{} supports a variety of storage formats such as row and column store with an optional support for compression. 
The data in a table is horizontally partitioned in small independent storage blocks. 
The size of each storage block is fixed, yet configurable.
The intermediate output of relational operators (e.g. filter) is stored in temporary output blocks, which follow a similar design as the storage blocks of the base tables. 

Each relational operator \wo{} has a unique set of input, described based on the semantics of the operator.
For instance, a select \wo{}'s input consists of a storage block and a filter predicate.
A probe join hash table \wo{}'s input is made up of a pointer to the hash table and a probe input block.
A \wo{} execution involves reading the input(s), applying the relational operator logic on the input(s) and writing the output to a temporary block.\footnote{Output of majority of the operators is represented in the form of storage block, except when the output itself is a data structure like hash table e.g. in the case of a build hash operator, or hash-based aggregation operators.}

\sys{} maintains a thread-safe global pool of partially filled temporary storage blocks.
During a \wo{} execution, a worker thread \textit{checks out} a block from the pool,  writes the output of the \wo{} to the block, and returns the block to the pool at the end of the \wo{} execution. 
Therefore a block is used by atmost one operator \wo{} simultaneously. 
This approach has two benefits: 1) We maintain locality of output block when output gets written to it and 2) Reduced memory fragmentation due to the reuse of output memory blocks. 

\subsection{Unit of Transfer (UoT)}
As \sys{} is fundamentally built on a block-based storage architecture, the UoT used in \sys{} is also defined w.r.t blocks.
As described earlier, the output of a relational operator \wo{} is stored in temporary blocks.
As soon as a block is full, it may be deemed ready for data transfer, subject to the UoT value.
For a small UoT value, the scheduler receives a signal as soon as an intermediate output block is full, after which it dispatches a \wo{} for the consumer operator for execution.\footnote{Partially filled blocks are scheduled for data transfer at the end of the operator's execution.}

\textbf{Interplay between block size and UoT}: For a given block size, we consider two extreme values for UoT. 
The smallest UoT is a single block. 
As soon as a block is produced, we transfer it to the consumer. 
The largest UoT is the entire intermediate table. 
We wait for the entire table to be produced before transferring it to the consumer. 

\subsection{Data Transfer Mechanism and Scheduling}\label{ssec:system-scheduling}
\begin{figure}[t]
	\centering 
	\includegraphics{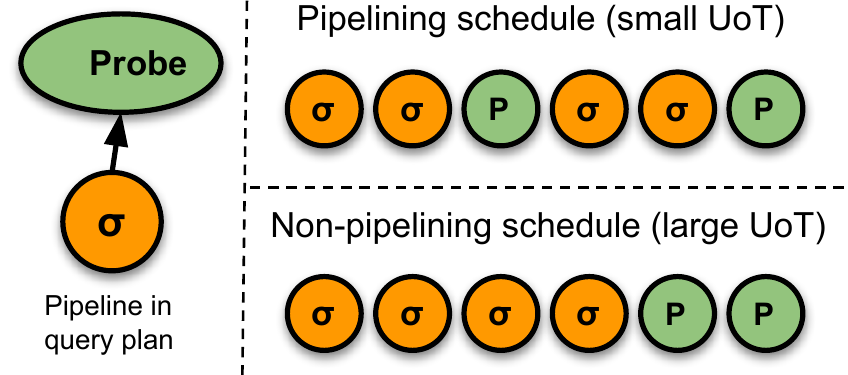}
	\caption{\textbf{Interplay between scheduling strategies and UoT values. A sample produce-consumer pair of a filter operator ($\sigma$) and a probe operator (P) for a hash join is shown on the left. On the right are two possible interleaving of the \wo{}s of these two operators, resulting in what are traditionally called `pipelining' and `non-pipelining' schedules}}
	\label{fig:pipelining-schedules}
\end{figure}

The scheduler for \sys{} supports different scheduling strategies.
A scheduling strategy impacts the sequence in which different \wo{}s for different operators are executed.
We view query processing at different values of UoTs as the outcome of two different scheduling strategies, as depicted in Figure~\ref{fig:pipelining-schedules}. 

For smaller UoT values, a consumer operator \wo{} is scheduled as soon as it is available.
At the higher end of the spectrum of UoT values, a consumer operator \wo{} is not scheduled until all the corresponding producer \wo{}s have finished execution. 

\sys{} scheduler allows development of sophisticated scheduling policies, such as implementation of an operator with an upper or lower limit on the number of concurrent consumer \wo{}s under execution, or executing operators under a specified memory budget.
\section{Discussion on Dimensions}\label{sec:dimensions}
In this section we identify dimensions that may have an impact on the performance of data transfer mechanisms for different values of UoTs.
We classify these dimensions into three categories: \textit{physical organization of data} (storage format and block size), \textit{execution environment} (parallelism and hardware characteristics), and \textit{structural aspects of query}.
We describe these dimensions below. 

\subsection{Block Size}
We first explain the concept of block size.
As the producer operator processes the input, it materializes the output to a temporary block.
The block size in \sys{} for a given table is fixed, and can be specified at the time of its creation.


We are interested in the impact of block size on the performance of the data transfer mechanisms. 
Consider data transfer between two operators: \texttt{select} operator $\rightarrow$ \texttt{probe} hash table operator.
A smaller output block size means that the block can potentially fill quickly. 
Therefore, compared to a larger block size, the \texttt{probe} \wo{} may be shorter. 

\subsection{Storage Format}
Data processing time is impacted by the way data is organized.
We look at two common storage formats: the row store format and the column store format.
In the column store format, values for a given column are stored in a contiguous memory region.
Scanning a single column results in a sequential memory access pattern, and generally good cache behavior.
In the row store format, all the columns of a tuple are stored in a contiguous region. 
Thus, scanning a particular column involves bringing unnecessary data (non-referenced columns) into the caches. Selecting all the columns in a row, however, is more efficient. 

\begin{figure*}[ht]
	\centering
	\begin{subfigure}[ht]{0.46\textwidth}
		\includegraphics{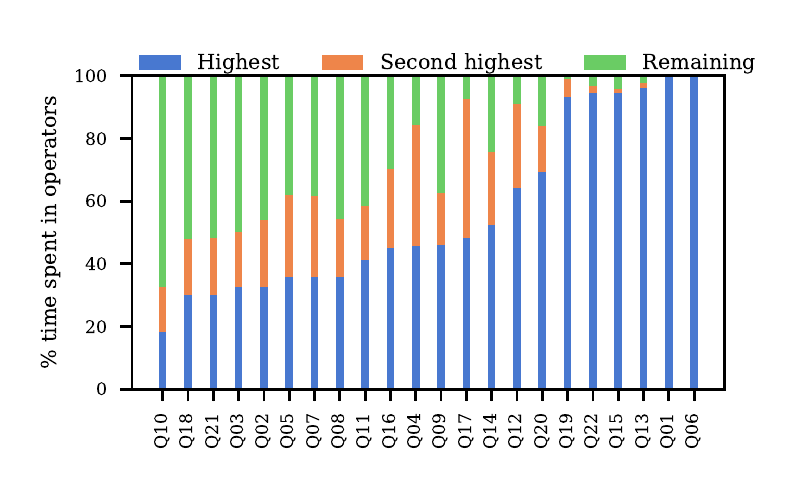}
		\caption{\textbf{Column store}}
		\label{fig:time-distribution-all-tpch-colstore}
	\end{subfigure}
	~
	\begin{subfigure}[ht]{0.46\textwidth}
		\includegraphics{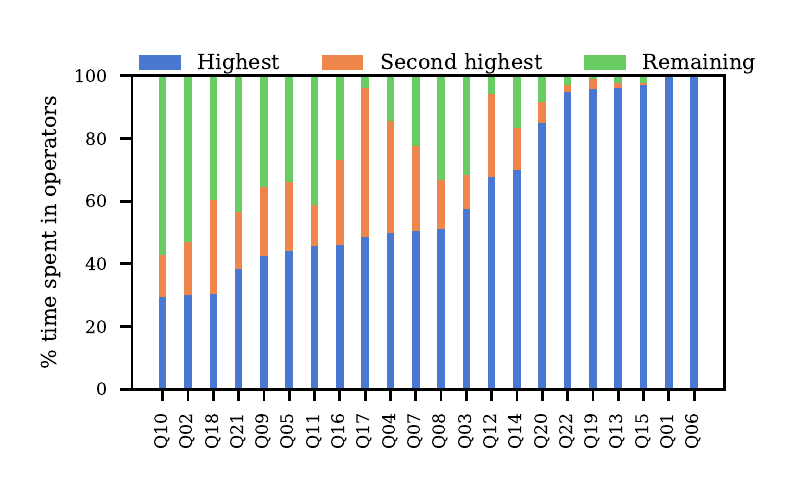}
		\caption{\textbf{Row store}}
		\label{fig:time-distribution-all-tpch-rowstore}
	\end{subfigure}
	\caption{\textbf{Distribution of time spent in each TPC-H (scale factor 50) query among its operators.}}
	\label{fig:tpch-operator-time-distribution}
\end{figure*}

Prior work has shown that column stores deliver better query performance for analytical workloads~\cite{DBLP:conf/sigmod/AbadiMH08}, especially for scan operators. 
Recent studies~\cite{quickstep-system} have shown that the performance gap between column stores and row stores is not as high as shown in previous work.
Therefore we explore both storage formats. 
For our comparison, we assume that all base tables are stored in the same storage format.
For micro-benchmarking in \sys{}, we note that the row store format is used for temporary tables irrespective of the storage format of the base tables.

\subsection{Parallelism}
We focus on two kinds of parallelisms in query processing: a) Inter-operator: processing multiple operators at the same time, and b) Intra-operator: parallel processing the work inside a single operator. 
Intra-operator parallelism is prevalent in modern database systems~\cite{monetdb, morsel, quickstep-system, vectorwise, DBLP:conf/sigmod/CruanesDG04, DBLP:journals/pvldb/RamanABCKKLLLLMMPSSSSZ13}.
Also note that, these two kinds of parallelisms can co-exist in a system~\cite{vectorwise, morsel, quickstep-system}. 
We would like to study the impact of intra-operator parallelism on the relative performance of data transfer mechanisms with different UoT values.

\subsubsection{Degree of parallelism}
The degree of parallelism (DOP) of an operator refers to the number of concurrent threads involved in executing \wo{}s of that operator. 
The \textit{scalability} of an operator (using $T$ threads) is its performance with DOP as $T$ relative to its performance when DOP is 1.

\subsubsection{Intra-operator parallelism in \sys{}}
For microbenchmarking, we use \sys{}. The scheduler of \sys{}, as discussed before, dispatches \wo{}s of relational operators to worker threads.
Thus the DOP of an operator at a given instance is the number of its \wo{}s under execution.
As in \sys{}, the number of \wo{}s of an operator can change over time, thus changing its DOP accordingly. 

\subsubsection{Interplay between DOP and UoT values}\label{sssec:dop-pipelining-interplay}
The UoT value used in query processing can have an impact on the DOP of the operators. 
Consider the example from Figure~\ref{fig:pipelining-schedules}.
We can observe that a smaller UoT value produces consumer \wo{}s less frequently compared to bigger UoT values.
Thus, the DOP of the consumer operator is small for small values of UoT. 
%

\subsubsection{Scalability}\label{sssec:scalability}
In theory, adding more CPU resources for an operator execution should offer linear speedup.
The assumption being that each parallel \wo{} operates at the same speed and thus by executing more \wo{}s concurrently, the overall execution time reduces proportionally. 

Linear speedups for operators (or for queries as a whole) are not always possible.
DeWitt and Gray~\cite{DBLP:journals/cacm/DeWittG92} propose reasons for less than ideal speedup for parallel databases such as startup costs, interference from concurrent execution and skew.
We can extend some of their ideas to in-memory systems.
For example interference can come from various sources such as contention due to latches, and shared use of a common bandwidth in a memory bus, or shared channels for data movement across NUMA sockets.

For an operator that exhibits poor scalability, increasing its DOP beyond a limit may degrade its performance.
Specifically, the execution time for each \wo{} of the operator may increase with a higher DOP value. 
We observe that poor speedup can have contrasting impact on performance depending on the UoT value chosen. It helps for smaller values of UoT, whereas at higher values of UoT performance is negatively impacted.
The reason for such contrasting behavior lies in the difference in the DOP values of a consumer operator in query processing for different values of UoT.
As discussed in Section~\ref{sssec:dop-pipelining-interplay}, the DOP of the consumer operator is lower for smaller values of UoT as compared to higher values of UoT.

\subsection{Hardware Prefetching}
We start with an explanation of hardware prefetching. 
It is a technique used by modern hardware to proactively fetch data in caches by speculating its access in the future.
The prefetcher observes patterns of data accesses from memory to caches and speculates the access of a data element in advance.
Prefetching hides the latency due to a cache miss and potentially improves performance.
There are two kinds of prefetching: spatial and temporal, and in this paper, we focus on spatial prefetching. 

Now we describe why prefetching is important to our study. 
Lower values of UoT generally results in a large number of context switches for \wo{} execution (c.f. Figure~\ref{fig:pipelining-schedules}).
Thus, having a lower value of UoT may affect the hardware prefetcher's ability to predict the data access patterns. 
Therefore, we are specifically interested in the impact of hardware prefetching at lower UoT values. 

In addition to the hardware-based prefetching implementation, there are software-based techniques for prefetching.
There is prior work on using software-based prefetching to improve the performance of relational operators~\cite{DBLP:conf/icde/ChenAGM04, rof}.
By focusing on hardware-based prefetching, we can observe the impact of the hardware prefetcher without modifying the implementation of the relational operators.

For our study, we run the queries with pipelining in two scenarios: a) when hardware prefetching is enabled (this is the default behavior of the hardware) b) when hardware prefetching is disabled (by setting bit 0 and 1 in Model-specific Register (MSR) at address 0x1A4) as per Intel's guidelines~\cite{intel-prefetching}.

\subsection{Query Plan Structure}\label{ssec:query-structure}
Complex queries like the ones in TPC-H contain several operators, and the impact of UoT values on overall query execution time is not immediately evident.

To analyze the impact of UoT values on the response time of a query, we conduct an experiment to dissect  the time distribution of the execution of TPC-H queries.
We focus on the most dominant operator (where the most of the execution time is spent) and the second most dominant operator for each query. 
Note that for this analysis, we run the queries with a high UoT value (the whole table) to avoid any overlap in time. 
The intuition is that if there is only one operator in the query where the majority of the query execution time is spent, small UoT values may not play a big role in the overall execution time of the query.

Figure~\ref{fig:time-distribution-all-tpch-colstore} shows the results of this experiment for base tables stored in a column store format. 
For some queries (Q1, Q6, Q13, Q14, Q15, Q19, Q22) the dominant operator takes up the majority of the query execution time (more than 50\%).
We also note that the dominant operator for many of these queries is a ``leaf'' operator (e.g. selection on a base table, building a hash table on a base table, aggregation on a base table). 
Therefore, depending on the query structure of a query, small UoT values may not provide significant advantage in improving the query execution time. 

In queries, sometimes, large data is pruned at the initial operators, e.g. due to a highly selective filter predicate or join condition, or due to application of sideways filters (e.g. LIP~\cite{DBLP:journals/pvldb/ZhuPSP17}). 
In such cases, very little data is passed on to the consumer operators, and consequently, the impact of low UoT values is not significant.

\section{Analytical Model}
\label{sec:model}

In this section we analytically model the performance difference for varying UoT values. The analytical model uses the dimensions introduced in the previous section such as number of threads used for execution (parallelism), UoT values, along with memory/cache access times and cache miss penalties (hardware prefetching). Our model is targeted towards in-memory environments, but it can be easily extended to other storage device settings, as we show in Section~\ref{ssec:model-disk}.
The model and analysis of memory usage differences will be presented in the next section.

The key idea that we exploit in our model is to focus on operations that result in a cost difference and to ignore common operations that occur irrespective of the UoT values. Many operations are common to query processing for all UoT values: e.g. the total cost of reading from L1 cache is the same irrespective of the schedule. 
As we are interested in the relative comparison of performance between two extreme values of UoT, it is safe to ignore the costs of operations that are common to both strategies, and focus on the additional work for each strategy that is not present in the other strategy. 

Additionally, we take into account the benefits of hardware prefetching when dealing with reading multi-megabyte blocks or UoTs in a sequential access pattern; the amortized cost of reading a block or UoT will be substantial lesser than when each block or UoT is read on its own without prefetching.
As the block or UoT is read into memory, initial few tuples likely incur an L3 cache miss, but we assume that the prefetcher can quickly detect the access pattern and thus beyond a point, the miss penalty will decrease. 

We analyze a basic producer-consumer pair, in which the producer is a select operator and the consumer is a probe operator for a hash-based join. This producer-consumer pair is commonly found in the query plans of TPC-H queries. For example, in the query plans for Q07 and Q19 from the TPC-H benchmark, selection is performed on the \textit{lineitem} table and the output is subsequently used to probe a join hash table, forming the \texttt{select}$\rightarrow$\texttt{probe} pair for data transfer. Table~\ref{table:analytical-model-notation} contains various parameters that we use to determine the costs for different scheduling strategies. We note here that the model and the results easily extend to multiple operators and/or different operators.

\begin{table}
	\begin{tabular}{|p{1.35cm}|p{6.2cm}|}
		\hline
		\textbf{Notation} & \textbf{Description} \\
		\hline
		$R_h$  & Cost of reading an UoT to memory hierarchy $h$ from a lower hierarchy h + 1 \\ \hline
		$AR_h$ & Amortized cost of reading an UoT sequentially to memory hierarchy $h$ from a lower hierarchy h + 1 \\ \hline
		$W_h$ & Cost of writing an UoT to memory hierarchy $h$ from a higher level hierarchy \\ \hline
		$IC$ & Cost of an instruction cache miss \\ \hline
		$M_h$ & Cost of missing a UoT at memory hierarchy h \\ \hline
		$N^{in}_{op}$ & Number of input UoTs for operator $op$ \\ \hline
		$N^{out}_{op}$ & Number of output UoTs for operator $op$\\ \hline
		$T$ & Number of threads in the system\\ \hline
		$B$ & UoT size \\ \hline	
	\end{tabular}
	\caption{\textbf{Notations used for the analytical model}}
	\label{table:analytical-model-notation}
\end{table}

For high UoT values equal to the size of the table, the output of the \texttt{select} operator is not immediately consumed by the \texttt{probe} operator; the \texttt{probe} operation is only initiated after the \texttt{select} operation is complete. Thus, writing the output of the \texttt{select} operation to memory, and reading the same UoTs as input to the \texttt{probe} operator is additional work done in the non-pipelining case. Additionally, an input probe UoT is likely to be cold in the caches when it is read for the \texttt{probe} operation.

Thus, for the case of high UoT values equal to the size of the table, the extra work done can be quantified as:
\begin{equation*}
W_{mem} \cdot N^{out}_{select} + AR_{L3} \cdot N^{in}_{probe} + p_1 \cdot N^{in}_{probe} \cdot  M_{L3}
\end{equation*} 

$W_{mem} \cdot N^{out}_{select}$ is the cost of writing the output of the \texttt{select} operator from cache to memory.

$AR_{L3} \cdot N^{in}_{probe}$ is the total cost of reading \texttt{probe} UoTs sequentially from memory, expressed as the amortized cost of reading a UoT sequentially times the number of UoTs. 

For the last term, note that a \texttt{probe} \wo{} has two input components: \texttt{probe} input UoT and a hash table.
As the reads in a hash table are random, it disrupts the sequential access pattern used for reading the \texttt{probe} input UoTs.
Therefore, we account for the cost in reading the \texttt{probe} input UoTs as $p_1 \cdot N^{in}_{probe} \cdot  M_{L3}$, where $p_1$ is the probability that there is a L3 cache miss for reading \texttt{probe} input after the context switch back from reading the hash table.

Next, we quantify the additional work done for the case of small UoT values.
The main conceptual difference in comparison to the execution for high UoT values is that the input for \texttt{probe} (which is the output of \texttt{select}) is presumed to be hot in caches while it is read to perform the probe operation. This leads to the following model:
\begin{align*}
&(N^{out}_{select}  +  N^{in}_{probe}) \cdot IC  + 
p_2 \cdot N^{in}_{probe} \cdot  (M_{L3} + R_{L3})  \\
& + p'_1 \cdot (M_{L3} + R_{L3} + W_{mem}) \cdot N^{in}_{probe} 
\end{align*}
Notice that for low UoT values, every \texttt{probe} \wo{} execution involves two context switches: First from \texttt{select} to \texttt{probe} and another from \texttt{probe} to \texttt{select}. 
Thus we account for two instruction cache misses; one for each of such context switch, which is represented by the term: $(N^{out}_{select} + N^{in}_{probe}) \cdot IC$.

Now, we explain the term $p_2 \cdot N^{in}_{probe} \cdot  M_{L3}$. It represents the cache misses due to disruption in sequential access pattern of \texttt{select} caused by intermittent \texttt{probe} operations.
The term $p_2$ is the probability of an L3 cache miss for the \texttt{select} operator after the context switch back from the \texttt{probe} operator. 

Finally, the term $p'_1 \cdot (M_{L3} + R_{L3} + W_{mem}) \cdot N^{in}_{probe}$ is analogous to L3 cache misses during probe input block reads in the non-pipelining case. Here we make the assumption that probe inputs are resident in L3. 
Thus, the probability of whether a probe input is read hot or not is dependent on the size of UoT. 

Due to factors such as reading in the relevant UoTs of the hash table for a probe operation and multiple threads sharing the L3 cache, each write for creating \texttt{probe} input, and subsequent \texttt{probe} input read may not be guaranteed to be served from the L3 cache; this gets exacerbated with higher values of UoT. So, we account for  the cost of reading and writing probe input UoTs.
The term $p'_1$ represents the likelihood that the reads and writes incur L3 cache misses, and is expressed as $min (1, 2B \cdot T/ size(L3))$.
The term $p'_1$ is smaller for small UoTs, and it is 1 for high values of UoTs and when $T$ is high.

\subsection{Quantifying the Difference}
\label{ssec:large-blocks-difference}

We now quantify the differences between the two extreme values of UoTs.
We first make a few observations below that will help simplify the analysis.
As large UoT values are typically a few megabytes, the instruction cache miss costs become negligible in this case. Thus, we can ignore the cost associated with instruction cache misses. 
Second, we observe that $N^{in}_{probe} = N^{out}_{select}$.
Thus, the ratio of costs of non-pipelining (informally large UoT) and pipelining (informally low UoT) strategies looks as follows:
\begin{equation*}
\resizebox{0.95\hsize}{!}{$\frac{W_{mem} \cdot N^{in}_{probe} + AR_{L3} \cdot N^{in}_{probe} + p_1 \cdot N^{in}_{probe} \cdot  M_{L3}}{p_2 \cdot N^{in}_{probe} \cdot  (M_{L3} + R_{L3}) + p'_1 \cdot (M_{L3} + R_{L3} + W_{mem}) \cdot N^{in}_{probe}}$}
\end{equation*}

This ratio can be simplified to 
\begin{equation*}
\frac{AR_{L3} + W_{mem} + p_1 \cdot M_{L3}}{p_2 \cdot (M_{L3} + R_{L3}) + p'_1 \cdot (M_{L3} + R_{L3} + W_{mem})}
\end{equation*}

Observe that $AR_{L3} \ll R_{L3}$, while both costs, $AR_{L3}$ and $R_{L3}$, are directly proportional to the size of UoT, $B$.
We consider the two representative cases of high and low UoTs values to estimate the difference between the two strategies.

\paragraph{\textbf{High UoT values}}
For high UoT values (size $> \frac{|L_3|}{2 \cdot T}$), $p'_1$ will be close to 1 and $p_2$ will be very low. Additionally, the cost contribution of $M_{L3}$ will be low in general, and $W_{mem}$ will be the dominant cost. We expect that $p_1 \cdot M_{L_3} \approx M_{L3} \cdot (p'_1 + p_2)$; $p_2 \cdot R_{L3} + p'_1 \cdot (R_{L3} + W_{mem}) \approx p'_1 \cdot (R_{L3} + W_{mem})$, which leads to $p'_1 \cdot (R_{L3} + W_{mem}) \approx AR_{L3} + W_{mem}$. Hence, the ratio given (1) will be very close to $1$.
Thus, we expect for high UoT values, the difference between two strategies to be negligible. 

\paragraph{\textbf{Low UoT values}}
Smaller UoT values result in a large number of work orders, which incurs a large overhead in storage management. Some examples for such overhead include creation cost of several UoTs, maintaining references for UoTs present in-memory, synchronisation costs in the data structures for storage management, etc. So, in this scenario, $p_2$ will be close to $1$, and $p'_1$ will have a lower value, though not negligible. The cost contributions from terms $AR_{L3}$, $p_2 \cdot M_{L3}$, $p_1 \cdot M_{L3}$, and $p'_1 \cdot M_{L3}$ will not be significant and we would expect that $W_{mem} \approx p_2 \cdot R_{L3} + p'_1 \cdot (R_{L3} + W_{mem})$. The ratio will be very close to $1$; since the cost of $W_{mem}$ is dominant, the cost of $p_2 \cdot R_{L3} + p'_1 \cdot (R_{L3} + W_{mem})$ can be slightly lower than $W_{mem}$, giving the execution with lower UoT values a slight advantage.


\subsection{Applying Model to Other Storage Settings}\label{ssec:model-disk}
Our model can be easily applied to other settings, such as storing data in a persistent store (such as SSD and hard disk drives) with a in-memory buffer pool. 
We change the parameters from Table~\ref{table:analytical-model-notation} appropriately to fit the persistent store setting.
The terms $p_1$ and $p_2$ can be nearly 0, assuming that the hash table is always kept in the buffer pool. 
Thus, the additional work done for large values of UoT is:
\begin{equation*}
R_{store} \cdot N^{in}_{probe} +  w_{store} \cdot N^{out}_{select}
\end{equation*}
which could be in the order of seconds for thousands of UoTs. 
The additional work done for lower UoT values is:
\begin{equation*}
N^{out}_{select} \cdot IC + N^{in}_{probe} \cdot IC
\end{equation*}
Note that this value is substantially lower (order of nanoseconds or microseconds for thousands of blocks) than that in the non-pipelining case.
Thus, the analytical model is consistent with the expected behavior for perstitent store-based systems. 

\section{Memory Characteristics}\label{sec:memory}
We now discuss the memory footprint for different UoT values.
We first formulate the memory footprint of two extreme UoT values (very low and very high -- equal to the table) individually, and then compare their memory behavior with each other. 

\begin{figure}[t]
	\centering 
	\includegraphics{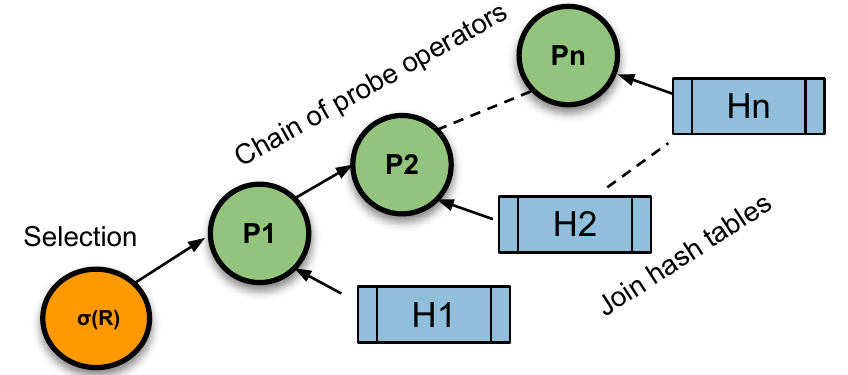}
	\caption{\textbf{A left-deep query plan fragment that shows a cascade of selection and multiple probe operators}}
	\label{fig:left-deep-plan}
\end{figure}

As an example, let us consider a cascade of selection and multiple probe operators as shown in Figure~\ref{fig:left-deep-plan}. (This pattern in common in many workloads including TPC-H and SSB.)

For low UoT values, once we read a tuple, it is processed by the selection operation first and if filtered, it is further processed by all the subsequent probe operators (subject to the join condition).
This means that all hash tables have to be constructed before the execution of selection-probe operators can begin. 

The case of high UoT value, where the value is equal to the size of the table, execution can be described as ``one join at a time".
The selection operation is completed first, followed by building of the hash table and then the probe. 
This means that only one hash table needs to be created at any point of time.
However this case of execution materializes the result of the selection (and successive probe operations).

We contrast the memory requirement for the leaf level join tree in Table~\ref{table:memory-comparison}.
We denote the size of the $i^{th}$ join hash table by $|H_i|$.
The size of the selection output is denoted by $|\sigma(R)|$ where $R$ is the input table.

\begin{table}[t]
\begin{tabular}{|c|c|c|}
\hline
\multirow{2}{*}{\textbf{Strategy}} & \multicolumn{2}{c|}{\textbf{Memory footprint}} \\ \cline{2-3} 
 & \multicolumn{1}{l|}{\textbf{Hash table}} & \multicolumn{1}{l|}{\textbf{Intermediate table}} \\ \hline
Low UoT value & $\sum_{i=1}^{n} |H_{i}|$ & 0 \\ \hline
High UoT value & $|H_1|$ & $|\sigma(R)|$ \\ \hline
\end{tabular}
\vspace*{1ex}
\caption{\textbf{Comparison of memory footprint for low and high UoT values}}
\label{table:memory-comparison}
\end{table}

We disregard the common elements contributing to the memory footprint to determine the difference of memory footprints such as current join hash tables, base tables, final join output.
Note that our analysis is done on the leaf level join, however it can be extended to any intermediate join easily.
Therefore the memory overhead comparison for the two strategies looks as below:
$$\text{Low UoT values:} \quad \sum_{i=2}^{n} |H_{i}|$$
$$\text{High UoT values:} \quad \sigma(R) $$

\subsection{Memory Overhead for high UoT values}
\begin{figure}[t]
	\centering 
	\includegraphics{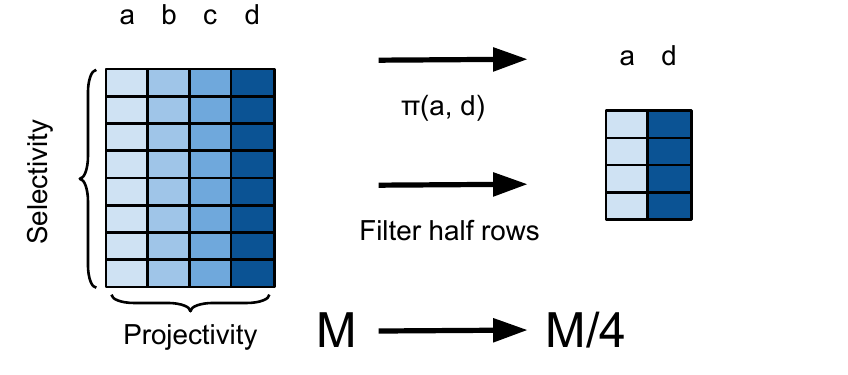}
	\caption{\textbf{Description of how memory footprint is reduced in a selection operation. Both selectivity and projectivity are 0.5. If base table's memory is $M$, output size is $M/4$}}
	\label{fig:selection-memory-reduction}
\end{figure}

We now dig deeper into the memory overhead caused by having a high UoT value, equal to the size of the table or intermediate result. The key relationship is between the size of the base table and the size of the materialized intermediate table. 
Typically a selection operation on a base table causes reduction in memory in two ways, as shown in Figure~\ref{fig:selection-memory-reduction}.
The first and the obvious aspect is selectivity of the filter predicate. 
We define selectivity as $s = N_s / N$, where $N_s$ is the number of rows that pass the filter predicate and $N$ is the number of rows in the input table. 
The other aspect is projectivity, which we define as $p = C_s / C$, where $C_s$ is the total size of the columns projected per tuple and $C$ is the total size of the columns in the base table per tuple.
We compute selectivity and projectivity relative to the size of the input table. 

In Figure~\ref{fig:selection-memory-reduction}, columns $a$ and $d$ are projected. 
We can see that $s = 0.5$, $p = 0.5$ and the size of the output is 25\% of the base table. 

\begin{table}[b]
        \centering
        \small
        \setlength\tabcolsep{2pt}
\begin{tabular}{|l|l|l|l|}
\hline
\textbf{Query} & \textbf{Selectivity (\%)} & \textbf{Projectivity (\%)} & \textbf{Total (\%)} \\ \hline
03 & 53.9  & 13.1  & 7.0  \\ \hline
07 & 30.4  & 18.3  & 5.6  \\ \hline
10 & 24.7  & 13.1  & 3.2  \\ \hline
19 & 2.1  & 13.1  & 0.3  \\ \hline
Average & 27.8  & 14.4  & 4.0  \\ \hline
\end{tabular}
\vspace*{0.5em}
\caption{\textbf{Memory reduction with input table lineitem}}
\label{table:memory-lineitem}
\vspace{0.5em}
\begin{tabular}{|l|l|l|l|}
\hline
\textbf{Query} & \textbf{Selectivity (\%)} & \textbf{Projectivity (\%)} & \textbf{Total (\%)} \\ \hline
03 & 48.6  & 8.7  & 4.2  \\ \hline
04 & 3.8  & 10.9  & 0.4  \\ \hline
05 & 15.2  & 5.8  & 0.9  \\ \hline
08 & 30.4  & 11.6  & 3.5  \\ \hline
10 & 3.8  & 5.8  & 0.2  \\ \hline
21 & 48.7  & 2.9  & 1.4  \\ \hline
Average & 25.1  & 7.6  & 1.8  \\ \hline
\end{tabular}
\vspace*{0.5em}
\caption{\textbf{Memory reduction with input table orders}}
\label{table:memory-orders}
\end{table}

\subsection{Memory Overhead for low UoT values}
As described earlier, the memory overhead for low UoT values is the combined memory of the hash tables that can be probed (except the current join).

Let us consider a single hash table.
Typically hash tables have fix sized buckets, say $c$ bytes.
Hash tables also have a load factor, say $f$ where $f \in (0, 1]$.
If $f$ is 0.5, the hash table gets resized as soon as its memory occupancy reaches 50\%. 
Therefore, the memory footprint of each entry that is inserted in the hash table is $c/f$.
If the input tuple has a size of $w$ bytes, and the input table's size is $M$ bytes, the resulting hash table size is $(M/w) \cdot (c/f)$.
For low UoT values, computing the memory overhead involves summing the hash table sizes for all the hash tables involved.

\paragraph*{\textbf{What UoT values have a lower memory footprint?}}
We have established the memory overhead for both low and high UoT values.
A natural question that would be of interest to system designers as well as their users to consider is: which UoT value results in lower memory overheads?
The answer is completely dependent on the query and its input tables characteristics. 
On the one hand, we see many cases when lower UoT values have a lower memory footprint than high UoT values, especially in Star Schema Benchmark (SSB)~\cite{DBLP:conf/tpctc/ONeilOCR09} queries which have small hash join hash tables. 
On the other hand, we show in the next section that sometimes high UoT values can also result in significantly low memory overheads. 

\subsection{Memory Analysis for TPC-H Queries}

We report the selectivity and projectivity values for selected TPC-H queries which contain a selection and probe pipeline of operators when the base table is \texttt{lineitem} or \texttt{orders}; the two largest tables in TPC-H schema. 

Table~\ref{table:memory-lineitem} and Table~\ref{table:memory-orders} presents the selectivity, projectivity and overall memory footprint of selection operations in selected TPC-H queries. 
A key takeaway from this figure is that even though the selectivity is high, due to the projectivity, the relative memory overhead of a select operation can be quite low.
In a star-schema or a snowflake-schema typically fact tables have large number of rows as well as large number of columns.
If few columns are projected from the fact table during a selection operation, the relative memory overhead can be low. 
Note that both selectivity and projectivity numbers reported above are without any optimization, thus they are on the higher side.
In practice there are many techniques to reduce both selectivity and projectivity. 

\paragraph*{\textbf{Techniques to lower selectivity}} 
Aggressive pruning techniques like Lookahead Information Passing (LIP) filters~\cite{DBLP:journals/pvldb/ZhuPSP17}, can substantially bring down the selectivity, sometimes by an order of magnitude. 
Query optimizers often push down predicates which can also help reduce the selectivity.

\paragraph*{\textbf{Techniques to lower projectivity}}
There are certain techniques to lower the projectivity, e.g. trading memory with computation. 
Consider the expression \texttt{l\_extendedprice * (1 - l\_discount)} which is found in many TPC-H queries.
A lazy evaluation of this expression implies projecting two attributes \texttt{l\_extendedprice} and \texttt{l\_discount}. 
However if the evaluation is folded with the selection operation, we can project only one attribute, which is the resultant expression. 

To compare the memory overhead between low and high UoT values, consider TPC-H Q07.
This query has a selection operation on \texttt{lineitem} followed by a cascade of three probe operations. 
Of the three hash tables required in this data transfer cascade, the second hash table is built on the entire \texttt{orders} table, which is around 2.4 GB in \sys{}.
The intermediate output of the selection operation is 2.8 GB without any optimization and 224 MB with bloom filter based pruning~\cite{DBLP:journals/pvldb/ZhuPSP17}. 
Therefore we can see that sometimes the memory overhead caused by lower UoT values can be substantially higher compared to the memory overhead caused by higher UoT values. 

We considered focusing on memory bandwidth utilization for different UoT values.
However we found that the bandwidth utilization during query execution in \sys{} was nowhere close to saturation. 
Thus we omit that discussion.
\section{Experimental Evaluation}\label{sec:experiments}
We now apply the proposed model on \sys{} and study the implication of UoT values on its performance.
Our goal is to understand the performance characteristics of queries for different UoT values and while doing so, to observe the impact of the various dimensions discussed in Section~\ref{sec:dimensions}. 

\subsection{Experimental Setup}\label{ssec:hardware-description}
We describe the hardware configuration, \sys{} specifications\footnote{We also experimented with block size as 512 KB, however we omit the results in the interest of space. The trends remain fairly similar to what we report.} and data set used for our evaluation in Table~\ref{table:hardware}.

\begin{table}[]
	\centering
	\begin{tabular}{|p{0.62in}|p{2.4in}|}
		\hline
		\textbf{Parameter} & \textbf{Description} \\ \hline
		Processor & 2 Intel Xeon Intel E5-2660 v3 2.60 GHz (Haswell EP) processors\\ \hline
		Cores & Per socket -- 10 physical and 20 hyper-threading contexts \\ \hline
		Memory & 80 GB per NUMA socket, 160 GB total \\ \hline
		Caches & L3: 25 MB, L2: 256 KB, L1 (both instruction and data): 32 KB \\ \hline
		OS & Ubuntu 14.04.1 LTS \\ \hline
		Data set& TPC-H benchmark~\cite{tpc-h} data (and queries) for scale factor 50 \\ \hline
		Block sizes &  128 KB, which is half of the per-core and private L2 cache size, and 2 MB, which comfortably fits in L3 cache. \\ \hline
		UoT values &  Low UoT is the same as block size and high UoT is the same as full table's size. \\ \hline
	\end{tabular}
	\caption{\textbf{Evaluation platform}}
	\label{table:hardware}
\end{table}

Our analysis focuses on performance of single socket, and thus we use only one of the two NUMA sockets on the machine.
Unless specified, we report the results for column store storage format and use all 20 threads as \sys{} workers.
The buffer pool size of \sys{} is configured with 80\% of the system's memory (126 GB).
We run each query 10 times and report the mean of the best three runs. 

\subsection{Results}
Now we present the results of our experimental evaluation. 
First we analyze performance of singleton operators, then progress to analyzing execution time of a bunch of operators together and finally study the execution time of the query.

\begin{figure*}[t]
	\centering
	\begin{subfigure}[ht]{0.46\linewidth}
		\includegraphics[width=\linewidth]{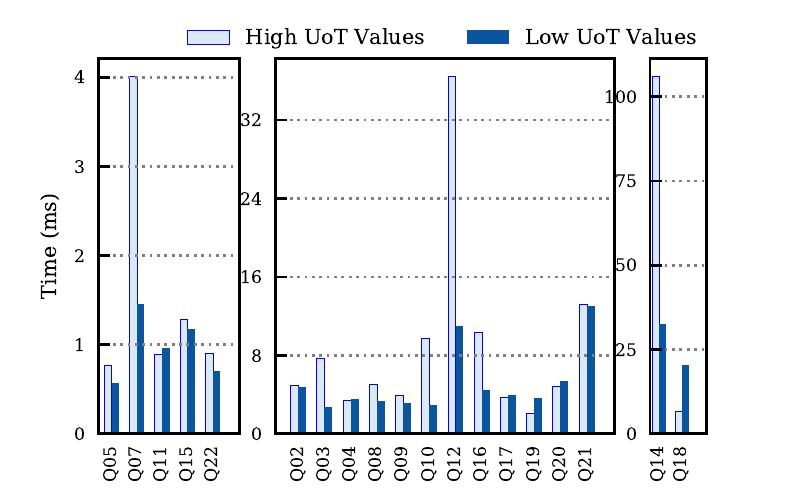}	
		\caption{Block size 128 KB}
	\end{subfigure}
	~
	~
	\begin{subfigure}[ht]{0.46\linewidth}
		\includegraphics[width=\linewidth]{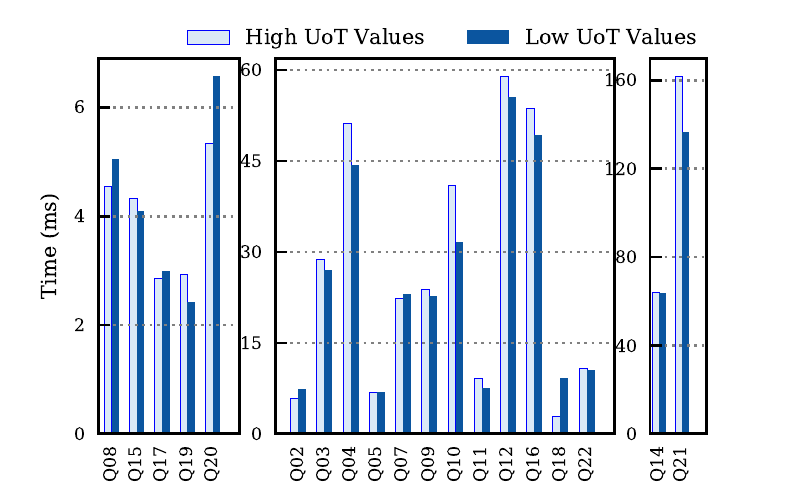}	
		\caption{Block size 2 MB}
	\end{subfigure}
	\caption{\textbf{Performance (per-task execution time) comparison of probe hash operator when it is the first consumer operator in a pipeline}}
	\label{fig:first-consumer-comparison}
\end{figure*}

\begin{figure*}[htb]
	\centering
	\begin{subfigure}[ht]{0.5\linewidth}
		\includegraphics[width=\linewidth]{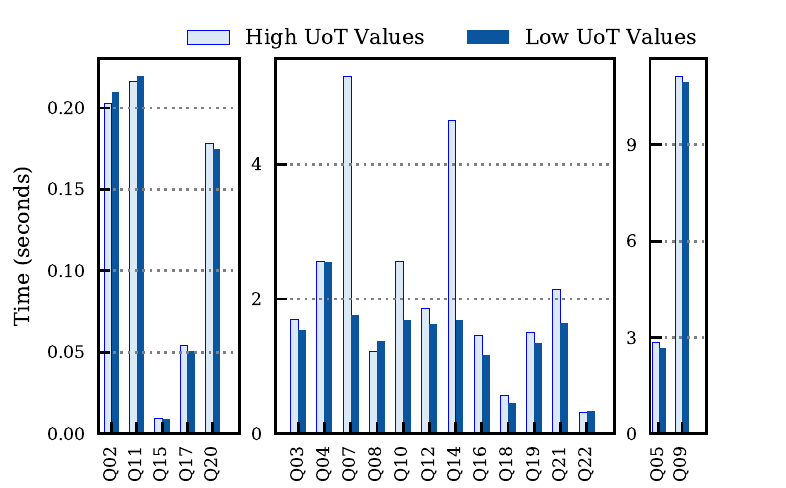}	
		\caption{Block size 128 KB}
	\end{subfigure}
	~
	~
	\begin{subfigure}[ht]{0.46\linewidth}
		\includegraphics[width=\linewidth]{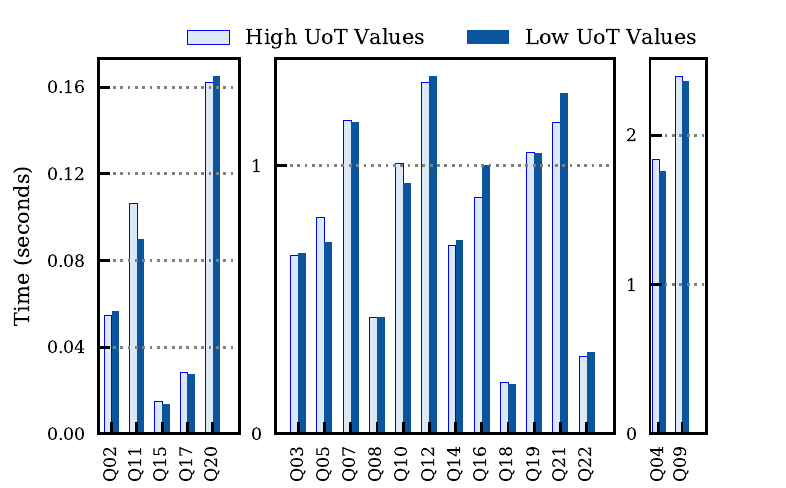}	
		\caption{Block size 2 MB}
	\end{subfigure}
	\caption{\textbf{Execution times of operator pipelines}}
	\label{fig:pipeline-comparison}
\end{figure*}

\begin{figure*}[]
	\centering
	\begin{subfigure}[ht]{0.46\linewidth}
		\includegraphics[width=\linewidth]{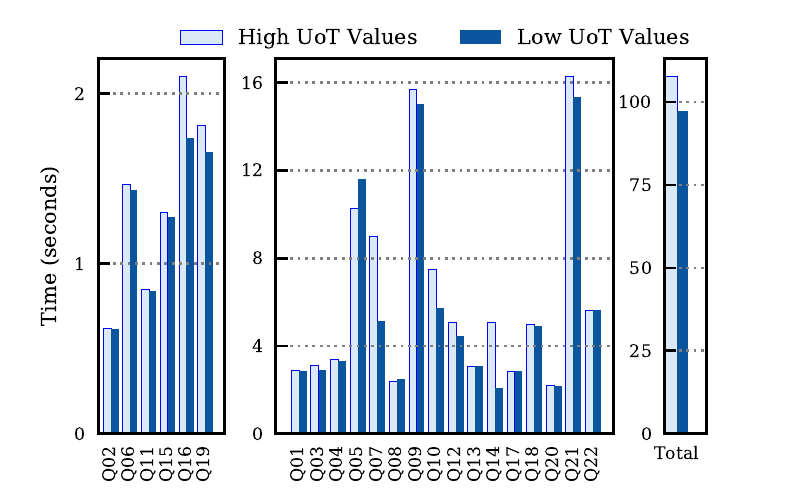}	
		\caption{Block size 128 KB}
	\end{subfigure}
	~
	~
	\begin{subfigure}[ht]{0.46\linewidth}
		\includegraphics[width=\linewidth]{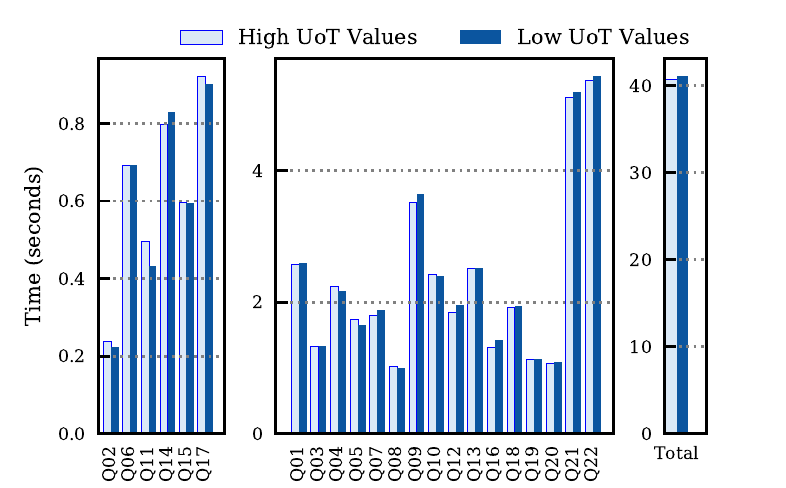}	
		\caption{Block size 2 MB}		
		\label{fig:colstore-2mb-query-time}
	\end{subfigure}
	\caption{\textbf{Execution times of queries}}
	\label{fig:query-comparison}
\end{figure*}

\begin{figure}[ht]
	\centering 
	\includegraphics{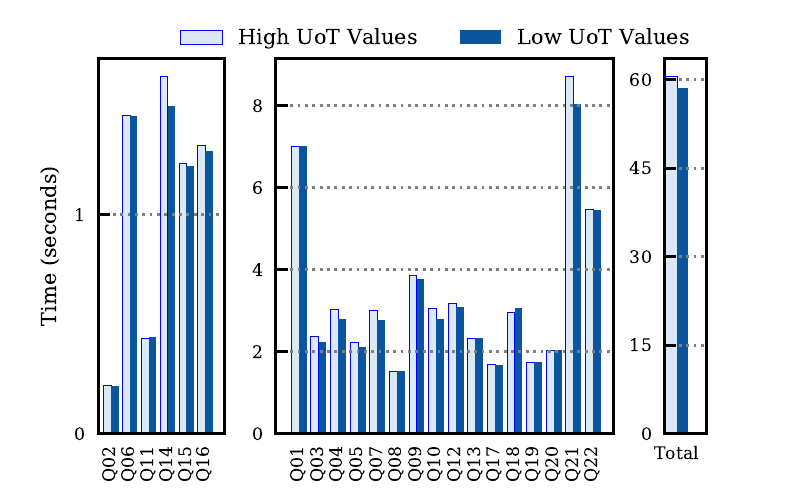}
	\caption{\textbf{Execution times of TPC-H queries for row store format and block size of 2 MB}}
	\label{fig:absolute-times-all-tpch-bs2mb-20threads-rowstore}
\end{figure}

\subsubsection{Performance of Consumer Operator}\label{ssec:consumer-time}
Low UoT values ensure that the consumer operator's input is hot in caches.
Does the hotness of the input in caches improve the performance of the consumer operator?
To find out we perform the following experiment.

\begin{figure}[ht]
	\centering
	\includegraphics[scale=2]{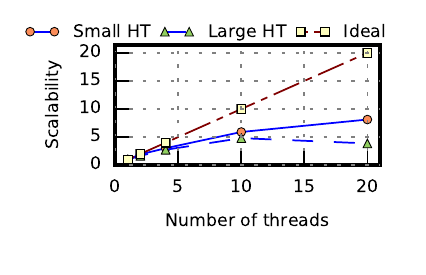}
	\vspace{-2em}
	\caption{\textbf{Scalability of two different probe operators in TPC-H Q07, compared against the ideal scalability. One probes a small hash table and the other probes a large hash table}}
	\label{fig:scalability-tpch-q07}
\end{figure} 

We focus on deep operator chains (select$\rightarrow$probe, as shown in Figure~\ref{fig:left-deep-plan}) from TPC-H queries.
Select$\rightarrow$probe operator chain is commonly found in OLAP queries. 
To break ties, we pick the chain with more \wo{} at its beginning. 

\begin{figure*}[]
	\centering
	\begin{subfigure}[ht]{0.46\linewidth}
		\includegraphics[width=\linewidth]{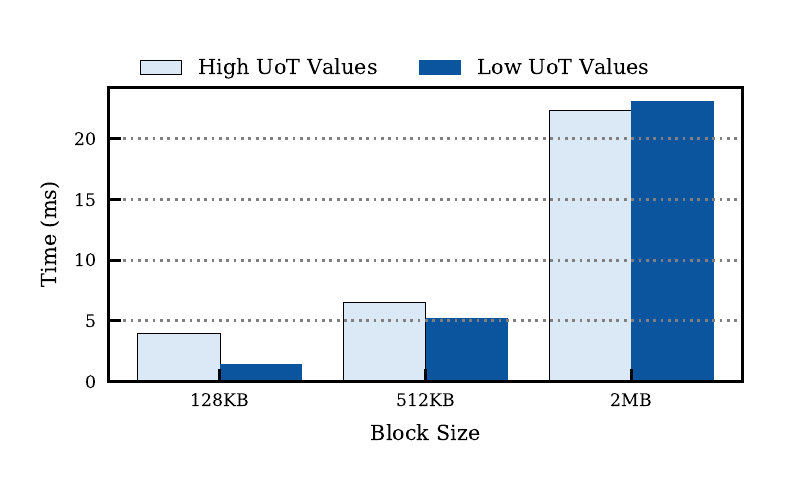}	
		\caption{Performance (per-task execution time) of the probe operator with better scalability}		
		\label{fig:more-scalable-probe}
	\end{subfigure} ~
	\begin{subfigure}[ht]{0.46\linewidth}
		\includegraphics[width=\linewidth]{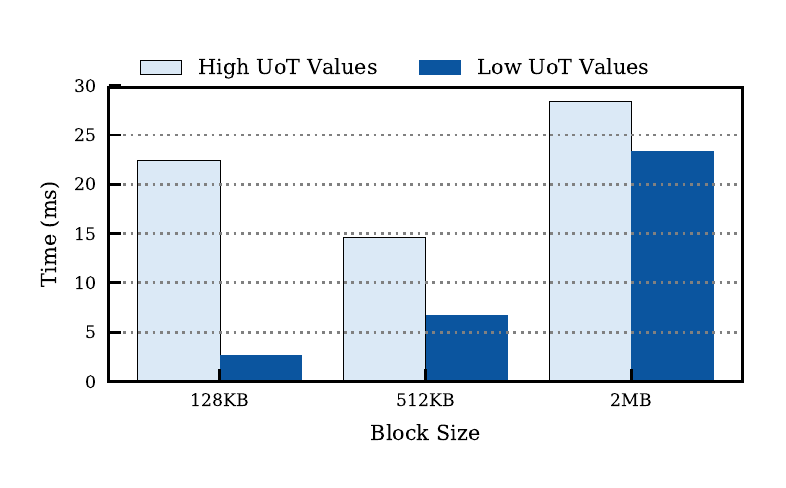}	
		\caption{Performance (per-task execution time) of the probe operator with poor scalability}		
		\label{fig:less-scalable-probe}
	\end{subfigure}	
		\caption{\textbf{Interaction of various dimensions that results in different scalability behaviors}}
	\label{fig:scalability-plots}
\end{figure*}

Figure~\ref{fig:first-consumer-comparison} plots the \wo{} execution times for the first consumer operator.
We can observe in Figure~\ref{fig:first-consumer-comparison} that a low UoT value generally benefits the performance of the probe operator. 
The extent of improvement diminishes as we increase the block size from 128 KB to 2 MB.
This behavior is consistent with the findings from the analytical model (c.f. Section~\ref{sec:model}).

\subsubsection{Performance of Operator Chains}\label{ssec:pipeline-time}
Having looked at the performance of the consumer operators, we zoom out to look at the performance of the complete chain of operators in each query.
The execution time of a chain of operators is the time elapsed between beginning and completion of its execution. 

Figure~\ref{fig:pipeline-comparison} shows the results of this experiment. 
For smaller block size, low UoT seems to outperform high UoT values only in some queries.
These queries are the ones in which low UoT has a superior operator performance (see Figure~\ref{fig:first-consumer-comparison}).
At 2 MB block size, operator chains from all queries perform equally well with both UoT values.

Notice that despite low UoT values having an edge for consumer operator performance, the extent of improvement does not match in the operator chain performance.
This is because typically a chain has other operators (e.g. producer) which often dominate its execution time. 

\subsubsection{Overall Execution Time:}\label{ssec:overall-execution-time}
After analyzing the performance of deep operator chains, we further zoom out to the execution times of the queries. 
Figure~\ref{fig:query-comparison} shows the results for query execution times for different UoT values. 
We can observe that low UoT values perform slightly better for smaller block sizes. 
As the block size increases, there is little difference between the two alternatives.
From these experiments we can conclude that though having low UoT values benefit individual operators, their impact on overall query performance is insignificant.

An alert reader may have noticed that query performance improves as the block size improves for both strategies.
It is due to \sys{}'s design which favors large multi-megabyte blocks. 
For smaller blocks the storage management and \wo{} scheduling overhead becomes higher. 
However we'd like to clarify that this is an orthogonal issue and doesn't affect our observations, also supported by the fact that the performance impact of block size variation is similar for both ends of the spectrum on UoT values.

\subsubsection{Effect of Storage Format}
Next, we study the effect of storage format of base tables on the performance of using low UoT values. 
We use two configurations, each with block size as 2 MB and all tables either stored in a) column store format or b) row store format. 
Note that in both configurations, temporary tables are stored in row store format. 

\paragraph*{\textbf{Performance comparison between pipelining strategies with row store}}
Recall the trends we discussed for performances of consumer operator (Section~\ref{ssec:consumer-time}), operator chain (Section~\ref{ssec:pipeline-time}) and entire query (Section~\ref{ssec:overall-execution-time}) for column store.
We observe similar trends with row store. 
In the interest of space we present one graph for query execution time using 2 MB block size in Figure~\ref{fig:absolute-times-all-tpch-bs2mb-20threads-rowstore}.

When we compare Figure~\ref{fig:absolute-times-all-tpch-bs2mb-20threads-rowstore} (row store) with Figure~\ref{fig:colstore-2mb-query-time} we notice two things.
First, the query performance is unaffected by the choice of UoT value.
Second, queries perform better with tables stored in column store. 
One of the key reasons why scans on row stores are slower than scans on column store is that processing a tuple involves fetching a lot of unnecessary data.
In column stores, we only process the necessary data from an attribute and avoid bringing unnecessary data in the caches. 

\subsubsection{Effect of Parallelism}\label{sssec:parallelism-effect}
Next, we study how intra-operator parallelism impacts performance with the two UoT values we have considered.
In this experiment we also report results from running \sys{} with a block size of 512 KB.

We present a result from TPC-H Q07 that shows the impact of scalability of operators on the performance.
We pick two probe operators from Q07 which are part of a single operator chain, one has a poor scalability and the other has slightly better scalability. 
The reasons for the poor scalability of the probe operator are a) It probes a large hash table b) The large hash table also brings about contention issues in \sys{}'s storage management subsystem.
We present the scalability of these probe operators in Figure~\ref{fig:scalability-tpch-q07}.

Now we want to know: how does the choice of a UoT value behave with these operators?
Figure~\ref{fig:more-scalable-probe} and Figure~\ref{fig:less-scalable-probe} present the performance of these two operators. 

First let us analyze the operator with the better scalability (whose input hash table is small). 
As the block size increases, both UoT alternatives keep up with the larger probe load and the per task execution time increases as expected. 

Now we discuss to the operator with poor scalability. 
Note that as block size increases from 128 KB to 512 KB, probe performance improves for high UoT value. 
As the block size increases, the contention on the storage management in \sys{} reduces and hence the performance improves. 
Going from 512KB to 2 MB, contention is no longer a dominant factor and hence the task execution time increases because of the added work in a larger block.

A question arises: Why don't we see similar behavior for low UoT values?
It is because by design, the degree of parallelism for low UoT values is smaller (as explained in Section~\ref{sssec:dop-pipelining-interplay}, and thus it is less prone to the contention discussed earlier. 

Therefore we can see that when using low UoT values, the system is more immune to scalability issues, as compared to the high UoT value alternative. 
Systems can have scalability issues due to various external (hardware interference, slow network) and internal factors (skew, poor implementation of operators).
We would like to stress that the reason for poor scalability of the probe operator in \sys{} when the build side is large, is not our focus\footnote{The specific probe operator in TPC-H Q07 can be made more scalable by partitioning the join or by putting the payload in the hash table bucket.}, rather it is the impact of poor scalability on the relative performance of the UoT alternatives.

\subsubsection{Effect of Hardware Prefetching}
We examine the effect of prefetching when using a low value of UoT.
For this experiment we use the tables stored in row store format. 
We run TPC-H queries with and without the hardware prefetcher. 
We note that the total workload execution time is only slightly (less than 10\%) better when prefetcher is enabled. 

In the row store format, to scan even a single attribute, lot of unnecessary data from the tuple is read.
As row store tuples are fixed width\footnote{Variable length attributes are stored in a separate region, with a pointer to the region stored in the tuple.}, the hardware prefetcher can detect the access pattern of scanning a single attribute. 

We are interested in understanding the impact of hardware prefetching on individual operators. 
We pick three operations from Q07 and compare their execution times with and without prefetching: selection, building a join hash table  and probing a join hash table.
Our results are presented in Table~\ref{table:prefetch-operator}.

\begin{table}[b]
\begin{tabular}{|l|l|l|l|l|l|l|}
\hline
\multirow{2}{*}{\textbf{Block size}} & \multicolumn{2}{c|}{\textbf{Select}} & \multicolumn{2}{c|}{\textbf{Build}} & \multicolumn{2}{c|}{\textbf{Probe}} \\ \cline{2-7} 
 & \multicolumn{1}{c|}{\textbf{Yes}} & \multicolumn{1}{c|}{\textbf{No}} & \multicolumn{1}{c|}{\textbf{Yes}} & \multicolumn{1}{c|}{\textbf{No}} & \multicolumn{1}{c|}{\textbf{Yes}} & \multicolumn{1}{c|}{\textbf{No}} \\ \hline
128KB & 0.06 & 0.08 & 2.0 & 1.9 & 0.8 & 0.8 \\ \hline
512KB & 0.2 & 0.3 & 8.5 & 7.6 & 2.2 & 0.9 \\ \hline
2MB & 1.1 & 1.5 & 38.0 & 32.7 & 3.9 & 3.1 \\ \hline
\end{tabular}
\caption{\textbf{Average task execution times in millisecond for the prefetching experiment. Yes (No) indicates that hardware prefetcher was enabled (disabled)}}
\label{table:prefetch-operator}
\end{table}

Our observations are as follows:
Prefetching generally benefits selection operator. 
This behavior is understandable as selection has a sequential access pattern and the prefetcher can recognize the strides in the memory accesses across tuples in the row store format.

Prefetching seems to worsen both probe and build hash performance in many settings. 
For both probe and build hash table operators, there are two data streams - a sequential access pattern for reading the input and a random read (probe) or random write (build) access pattern follows the initial read.
We suspect that due to a mix of these two access patterns, prefetching does not benefit these operators. 

We ran the same experiment for column store format and found little to no performance difference due to prefetching. 
We speculate that hardware prefetching does not make any significant contribution to an already optimized access pattern of column stores.

\subsection{Summary of Experiments}\label{sec:experiment-summary}
We described a large number of experiments in Section~\ref{sec:experiments}. 
In this section, we summarize our findings and connect them back to the dimensions described in Section~\ref{sec:dimensions}.
Recall that our focus is to understand the relative performance of the two UoT value alternatives which are at the extreme ends of the spectrum.
Our high level conclusion is that in the in-memory setting, for systems using block-based architecture, the performance is similar with these two alternatives.
We now discuss the impact of individual dimensions.

\paragraph*{\textbf{Block size}}
We find that a higher block size bridges the gap between the performance of the low UoT values and high UoT values. 
A larger block size results in a lower degree of parallelism for operators in a pipeline and thus also aides those operators that suffer from poor scalability. 
A very large block size however can cause memory fragmentation.
It may also result in a low DOP which could lead to underutilization of CPU.

\paragraph*{\textbf{Parallelism}}
Parallelism can affect the performance of the two UoT alternatives.
We saw in Section~\ref{sssec:parallelism-effect}, using a low UoT value can be more resilient to performance issues arising due to poor scalability.

\paragraph*{\textbf{Storage Format}}
The performance gap between the two UoT alternatives is largely unaffected by the choice of base tables' storage format.
We note that individually some queries execute faster when ran on tables in column store format. 
The benefit of using column store format over row store format is the highest for base tables (typically leaf operators in a query plan).
As the number of attributes in tables reduce from the leaf operators in a query plans to the root, the advantage of using a column store starts to diminish. 

\paragraph*{\textbf{Hardware Prefetching}}
Hardware prefetching improves the performance when using a low UoT value. 
The effect is more prominent in row store than in column store format. 
We saw that prefetching improves scan performance in a representative query.
Prefetching had an adverse effect on probe and build hash operators. 

As increasing L3 cache size becomes prohibitive due to power constraints, hardware prefetching techniques are gaining attraction~\cite{DBLP:journals/csur/Mittal16c}. 
Combined with the software-based prefetching efforts~\cite{DBLP:conf/icde/ChenAGM04, rof}, hardware prefetching could provide greater benefits in the future. 
\section{Conclusion and Future Work}\label{sec:conclusions}
In this paper, we make an observation that the boundary between a pipeline and block-based query execution is not well defined, and we introduce a clearer way to think about the data transfer mechanism between operators in a query processing pipeline. 

We discuss the many dimensions that impact the performance of ``pipelining'' and ``non-pipelining'' strategies for query processing.
We present an analytical model that estimates the performance of each strategy as a function of block size and cache misses. 
Our analytical model, as well as empirical evaluation on \sys{} database system shows that pipelining and non-pipelining strategies are not very different w.r.t. query performance. 
We demonstrate that the memory consumption of these two strategies could be quite similar in some cases; in fact sometimes non-pipelining can have a lower memory overhead, with the help of bloom-filter based pruning techniques.
We discuss the reasons for their similarities and dissimilarities throughout the paper. 

One key conclusion of the study is that the impact of pipelining on a query depends on the structure of the query plan and the execution time distribution of its operators.
Though pipelining speeds up some consumer operators due to better cache behavior, the improvements are not substantial when considered w.r.t. the total query execution time.
We showcase a surprising effect of scalability of operators on the performance of pipelining and non-pipelining strategies.

For the future work, we are interested in revisiting the assumptions made for pipelining in the context of new memory settings where the storage hierarchy likely looks far more complex than it does today.

\balance
\bibliographystyle{abbrv}
\bibliography{refs}

\end{document}